\documentclass[conference]{IEEEtran}
\IEEEoverridecommandlockouts
\usepackage{cite}
\usepackage{amsmath,amssymb,amsfonts}
\usepackage{algorithmic}
\usepackage{graphicx}
\usepackage{textcomp}
\usepackage{xcolor}
\usepackage{hyperref}
\usepackage[frozencache,cachedir=.]{minted}
\def\BibTeX{{\rm B\kern-.05em{\sc i\kern-.025em b}\kern-.08em
    T\kern-.1667em\lower.7ex\hbox{E}\kern-.125emX}}
\begin{document}

\title{A Universal System for OpenID Connect Sign-ins with Verifiable Credentials and Cross-Device Flow \\
}

\author{\IEEEauthorblockN{1\textsuperscript{st} Felix Hoops}
\IEEEauthorblockA{\textit{Department of Computer Science} \\
\textit{Technical University of Munich}\\
Munich, Germany \\
felix.hoops@tum.de}
\and
\IEEEauthorblockN{2\textsuperscript{nd} Florian Matthes}
\IEEEauthorblockA{\textit{Department of Computer Science} \\
\textit{Technical University of Munich}\\
Munich, Germany \\
matthes@tum.de}
}

\maketitle

\begin{abstract}
Self-Sovereign Identity (SSI), as a new and promising identity management paradigm, needs mechanisms that can ease a gradual transition of existing services and developers towards it. Systems that bridge the gap between SSI and established identity and access management have been proposed but still lack adoption. We argue that they are all some combination of too complex, locked into specific ecosystems, have no source code available, or are not sufficiently documented.
We propose a comparatively simple system that enables SSI-based sign-ins for services that support the widespread OpenID Connect or OAuth 2.0 protocols. Its handling of claims is highly configurable through a single policy and designed for cross-device authentication flows involving a smartphone identity wallet. For external interfaces, we solely rely on open standards, such as the recent OpenID for Verifiable Credentials standards.
We provide our implementation as open-source software intended for prototyping and as a reference. Also, we contribute a detailed technical discussion of our particular sign-in flow. To prove its feasibility, we have successfully tested it with existing software and realistic hardware.
\end{abstract}


\section{Introduction}
Identity management has long been and will likely stay a core pillar of our Internet service ecosystem and economy. From simple social media accounts to rigorous know-your-client (KYC) compliant financial service accounts, identity is everywhere. And while the principles of authentication and authorization seem eternal, their implementation evolves with the times. Isolated centralized solutions have largely given way to federated identity management with hints of user-centric identity, such as data sharing consent prompts after sign-ins\cite{schardongSelfSovereignIdentitySystematic2022}. In this age of the single-sign-on (SSO), a handful of large corporations act as Identity Providers (IdP) and keep user account data that any service can integrate, given the IdPs permission. This has created an alarming concentration of data, censorship potential, and, ultimately, power.


At the core of the issues is that the IdP takes an active role in authentication and authorization processes. Thus, it always knows what its users access and when. Denying service to individuals or groups of users\textemdash be it by design or by accident\textemdash is as easy as taking no action.
The concept of Self-Sovereign Identity (SSI) has evolved as the ultimate form of user-centric identity\cite{allen2016path}, reclaiming the active role for users. Initially described through a set of ten principles by Christopher Allen \cite{allen2016path}, SSI has become much more tangible through technical standards and open-source implementations.


Naturally, adopting and integrating an identity management approach fundamentally different from today's established Identity and Access Management (IAM) is a challenge. Even though there are initiatives from several sizable actors, including the European Union \footnote{\url{https://ec.europa.eu/digital-building-blocks/sites/display/EBSI/EBSI+Verifiable+Credentials}}, change will likely need to be gradual and more organic. Existing IAM systems need to be phased out over time, and users must adopt and understand the new paradigm.
Simple sign-in procedures arguably cover a majority of today's Internet use cases. Thus, building a system that can\textemdash at the very least temporarily\textemdash be used to bridge the gap between SSI credentials and established IAM solutions is necessary, and the few existing proposals have not seen significant adoption.

OpenID Connect (OIDC) \cite{openidFinalOpenID} is arguably one of the most widely used IAM protocols. General implementation support is good\footnote{e.g., \url{https://www.npmjs.com/package/oidc-client}} and implementations tailored to popular web frameworks are commonly available\footnote{e.g., \url{https://www.npmjs.com/package/oidc-react}, \url{https://www.npmjs.com/package/next-auth}, \url{https://www.npmjs.com/package/angular-oauth2-oidc}}, it is the protocol of choice for today's dominant IdPs\footnote{\url{https://developers.google.com/identity/openid-connect/openid-connect}}, and it is possible to integrate into arbitrarily complex IAM setups via Keycloak\footnote{\url{https://www.keycloak.org/docs/latest/server_admin/}}. This is likely why existing proposals for SSI bridges have mainly focused on bridging to OIDC\cite{kuperbergIntegrationSelfSovereignIdentity2022}, which we improve upon.


In this work, we present the status quo of SSI bridging and highlight remaining problems. We then propose a simple yet powerful SSI-to-OIDC bridge that service providers can deploy to adopt SSI-based sign-in while still using the familiar OIDC. As part of this, we advocate for the separation of issuer and relying party software. We design and implement such a bridge that we make available as free and open-source software intended for prototyping and as a reference\footnote{\url{https://github.com/GAIA-X4PLC-AAD/ssi-to-oidc-bridge}}. In the process, we emphasize cross-device protocol flows and the use of the recent standardization efforts belonging to the OpenID for Verifiable Credentials family \cite{openid4vcWhitepaper}. Also, we provide a deep technical discussion of one of these flows: the authorization code flow. Finally, we validate our bridge with the involvement of existing software, running everything on separate physical devices.

\section{Background}
In this section, we provide the minimum necessary background knowledge needed to follow this work. We provide an overview of the concept of Self-Sovereign Identity, introduce OpenID Connect with an emphasis on the core standard, and finally explore the newest additions to OpenID Connect that go beyond the core specification to provide some support for Self-Sovereign Identity.

\subsection{Self-Sovereign Identity}

SSI is a decentralized identity management paradigm that aims to return a higher degree of privacy and control to individual end-users \cite{allen2016path} by letting them hold their own user data. As of writing, SSI has gained momentum in research and policy communities, but large-scale adoption is yet to come.

Key to the technical implementation of SSI are the W3C Verifiable Credential (VC)\cite{w3VerifiableCredentials} and W3C Decentralized Identifier (DID)\cite{w3DecentralizedIdentifiers} standards. The VC standard defines a format for expressing claims, enabling the creation of digitally signed statements about subjects such as individuals, organizations, or machines. These credentials can be cryptographically verified.
Complementing VCs, the W3C DID standard establishes a pattern for creating and managing globally unique decentralized identifiers that represent the subjects and issuers of VCs. The actual implementation is up to each specific DID method, allowing for great flexibility. Usually, these work by leveraging asymmetric encryption keys, similar to how blockchain accounts work.

For end-users, the entry into any SSI ecosystem likely begins with a smartphone wallet application.
SSI wallets are comparable to crypto wallets in that they store private key material. However, their keys represent DIDs. In addition, SSI wallets also provide means to store VCs, display VCs, and implement protocols to present and receive VCs.

Presenting a VC happens in the form of a Verifiable Presentation (VP) that contains at least one VC and a challenge and is signed by the holder to prevent replay attacks. A relying party (RP) can cryptographically verify the VP and its VC(s) without interacting with the issuer. If designated in a VC, the issuer will also query a status list to ensure the VC is not revoked.




\subsection{OpenID Connect}

OpenID Connect (OIDC)\cite{openidFinalOpenID} is an open standard designed for secure user authentication and authorization. It is built on top of the OAuth 2.0\cite{rfc6749} authorization framework, providing an additional layer of identity verification. OIDC facilitates the exchange of user information between the identity provider (IdP) and the relying party (RP) in a secure and standardized manner.

The OIDC protocol introduces a set of standardized identity flows adapted from OAuth 2.0, such as the Authorization Code Flow, Implicit Flow, and Hybrid Flow, allowing clients operated by an RP to request and obtain identity information about users. To limit what information a client gets, OpenID Connect defines a set of scopes that determine the user attributes transmitted to the client during the sign-in process. Ultimately, the client ends up receiving tokens. An \texttt{id\_token} is used for identity data, such as an email address. An \texttt{access\_token} encodes information relevant to access control, such as membership in a user group.

One of the strengths of OpenID Connect is its ability to support single sign-on (SSO) scenarios, enabling users to log in once and access multiple services without the need for separate authentications.

\subsection{OpenID for Verifiable Credentials}

Recently, OpenID Connect has been expanded by OpenID for Verifiable Credentials (OID4VC)\cite{openid4vcWhitepaper}. Originally, OID4VC was a family of three specifications:

\begin{enumerate}
\setlength{\itemindent}{-1em}
    \item[] \textbf{OpenID for Verifiable Credential Issuance (OID4VCI)} defines an API for VC issuance from a server to a wallet. It essentially covers the download of a signed credential.
    \item[] \textbf{OpenID for Verifiable Presentations (OID4VP)} specifies how a wallet can present a VP to a server based on an OAuth 2.0 flow \cite{openidOpenIDVerifiable}.
    \item[] \textbf{Self-Issued OpenID Provider v2 (SIOPv2)} enables SSI wallets to act as OIDC Providers \cite{openidSelfIssuedOpenID}.
\end{enumerate}


Since then, more have been added\footnote{\url{https://openid.net/sg/openid4vc/specifications/}}, but considering our focus on sign-ins, OID4VP and SIOPv2 are the most relevant specifications of that family. They can be combined to create a standardized way for smartphone wallets to authenticate and authorize a user with claims from VCs.

\section{Related Work}
With the field of Self-Sovereign Identity (SSI) having enjoyed steady attention from researchers in the past years, several authors have contributed work related to the topic of integrating SSI into established IAM systems. These contributions roughly fit into three categories. We start by mentioning surveys about integrations of SSI into existing IAMs, then go over noteworthy integrations with OIDC, and finally, integrations with other IAMs.

A survey by Kuperberg et al. \cite{kuperbergIntegrationSelfSovereignIdentity2022} looking at SSI integrations for established IAM protocols identifies only seven relevant candidates. The majority are commercial, and only 2 of them are available as open-source software. They note a need for code examples and an explanation of implementation specifics.

In another survey, Grüner et al. \cite{grunerAnalyzingInteroperabilityPortability2021} formalize and compare different interoperability concepts for SSI. They are unable to clearly identify a superior concept. Also focused on interoperability, Yildiz et al. \cite{yildizTutorialInteroperabilitySelfsovereign2022} once more emphasize the need for true interoperability in SSI and describe it as a key requirement for further adoption. They introduce an SSI reference model that structures SSI components and functionalities into different layers. They also point out where specific protocols and standards could be applied on these layers, including OIDC technology standards.

A general survey of the state of SSI from 2022 by Schardong et al. \cite{schardongSelfSovereignIdentitySystematic2022} notes that protocol integration into established IAMs is vital to pave the adoption for SSI. The authors provide a comprehensive list of existing efforts.

Grüner et al. introduced a system facilitating IAM between the old SSI ecosystems of Jolocom and uPort, as well as the still-existent Hyperledger Aries, and the established IAMs OIDC and SAML2 over the course of two works \cite{grunerIntegrationArchitectureEnable2019, grunerATIBDesignEvaluation2021}. The presented system works as a two-way integration and consequently assumes the role of issuer and relying party simultaneously. An elaborate trust model, in combination with attribute mapping support, enables the forwarding of claims under unified names while marking untrusted claims by renaming them further. The most significant issue with this architecture lies in its complexity and the significant work needed to integrate new blockchain-based SSI ecosystems.

Lux et al. \cite{luxDistributedLedgerbasedAuthenticationDecentralized2020} propose an integration architecture that allows OIDC sign-in via SSI based on a blockchain public key infrastructure. Namely, they are looking at Sovrin, which uses Hyperledger Indy. In their implementation, they write attributes to the OIDC \texttt{id\_token} to avoid reliance on non-default OIDC features. Unfortunately, the solution is exclusive to Hyperledger Indy.

At least two open-source solutions without academic contribution exist. The first one is the Province of British Columbia's Verifiable Credential Authentication with OpenID Connect (VC-AuthN OIDC)\footnote{\url{https://github.com/bcgov/vc-authn-oidc}}. It supports a cross-device flow where the wallet is interfacing via DIDComm and uses custom presentation requests. Information from presented credentials is mapped into the OIDC \texttt{id\_token}. Unfortunately, the implementation seems to be mostly, if not only, compatible with their BC Mobile Wallet, and the trust model is unclear.

The second open-source effort identified is IdP Kit\footnote{\url{https://github.com/walt-id/waltid-idpkit}} from walt.id and looks promising. From what the documentation\footnote{\url{https://docs.walt.id/v/idpkit/getting-started/quick-start}} show, the project focuses on OIDC compliance and additionally offers sign-in based on Non-Fungible-Token possession. The biggest downsides seem to lie in the lack of detail regarding wallet interaction implementation and the amount of required configuration.
However, public deployments were unavailable, and recent builds were failing, as also acknowledged in the documentation.

Moving on to other IAMs, Hong et al. \cite{hongVaultPointBlockchainBasedSSI2020} develop an architecture that allows sign-in through OAuth 2.0, which is also used in OIDC. It is heavily blockchain-focused and requires one identification smart contract to be deployed per user. This is likely too complex for the average user in the near future, and leaves open the question of who will pay user registration costs. Since then, the standardization of DIDs has offered more usable options for mass adoption.

Yildiz et al. \cite{yildizConnectingSelfSovereignIdentity2021} present an architecture making SSI credentials available for SAML sign-in. The implementation relies on Hyperledger Indy and Aries. It supports a cross-device flow via DIDcomm. That means that the SSI-facing interface of the architecture is locked into Hyperledger Indy.

\section{Open Challenges}
In this section, we identify and examine open challenges in the area of SSI bridges. Our aim is not to highlight the general complexity of a bridge, but rather to focus on aspects that have not, or have not sufficiently, been addressed by other sources.

\subsection{Fragmented Presentation Protocols}
The exchange of a VP from a wallet to a server has long been fragmented in terms of protocols. Many wallet developers simply built their own. Also, most of these protocols lack standardized support for the cross-device flow that will likely dominate as wallets almost exclusively come in the form of mobile apps. Both of these are currently beginning to change, with the OID4VP and SIOPv2 standards gaining attention and adoption. The former standard features developer considerations that shed some light on cross-device flows \cite{openidOpenIDVerifiable}. However, the standard is still in development, and especially the hand-off back from phone to desktop is still too abstract. Contemporary wallet apps feature no mechanism to directly redirect a browser running on another device.



\subsection{SSI Ecosystem Dependence}
While interoperability is one of the fundamental ideas of SSI, implementers in practice have shown a tendency to build SSI ecosystems with a certain level of lock-in. This can easily happen through using DID methods living on a specific blockchain, a custom solution for trusted issuer management in the absence of a common standard, or custom status list implementations seeking more decentralization than StatusList2021\cite{w3StatusList} can provide. That leads to a bridge that is suitable only for a very select group of issuers and, by extension, service providers.

\subsection{High Bridge Complexity}
Previous bridging solutions tend to be fairly complex in design and setup, which goes entirely against the idea of simplifying SSI adoption. This can stem from supporting too many SSI ecosystems with custom drivers that increase the necessary amount of configuration, often also relying on multiple configuration interfaces. In practice, bridges are not just desirable for legacy services but also as an easy solution to prototype and build new ones. They need to be simple to run.

Furthermore, designs like the one by Grüner et al. \cite{grunerATIBDesignEvaluation2021} incur additional complexity by including full issuance capability when it arguably is not necessary for many use cases. Embracing SSI does not necessitate issuing every piece of user information. Only key attributes that conceivably have value outside the service of origin must be issued.


\subsection{Lack of Concrete Technical References}
As also noted by Kuperberg et al. \cite{kuperbergIntegrationSelfSovereignIdentity2022}, there is a distinct lack of code examples and practical implementation considerations. Specifications leave room for interpretation or significantly abstract parts of a flow. While seeing a full authorization code flow in less than ten steps is great as a gentle high-level introduction, it is not sufficient to base an implementation upon.

\section{System Architecture}
\begin{figure*}[tb]
\centering
\includegraphics[width=0.8\textwidth]{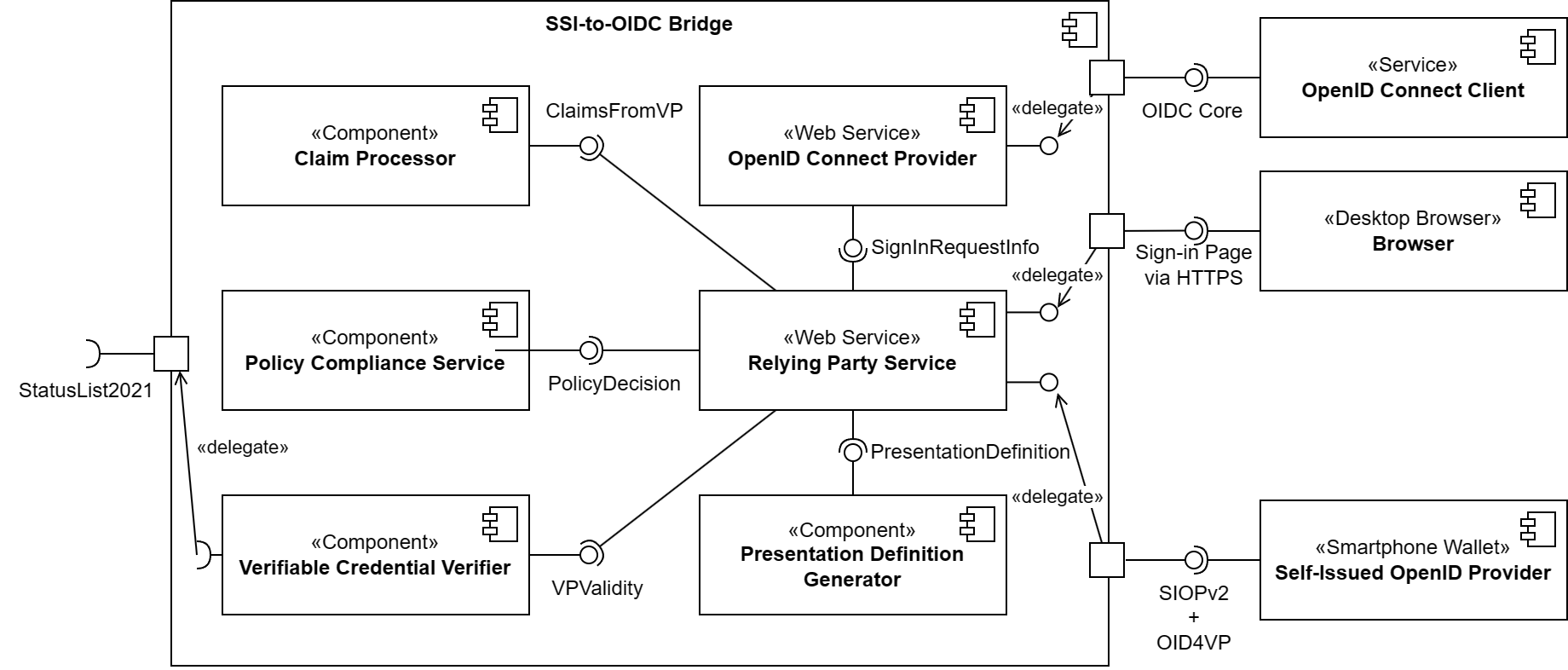}
    \caption{A component diagram of the SSI-to-OIDC Bridge's conceptual components and its interfaces.}
    \label{fig:architecture}
\end{figure*}

From a high-level perspective, the architecture we propose is two OIDC Providers with one nested inside the other. While the first one, which is realized through SIOPv2 with OID4VP, could, in theory, function as a standalone one, the reliance on the non-OIDC-Core \texttt{vp\_token} makes it unsuitable for use with most clients.
To remedy this, the second OIDC Provider encapsulates this first one and communicates entirely based on the OIDC Core that any current OIDC client supports.

In the remainder of this section, we discuss how that design translates into conceptual components and present, as well as explain, some further decisions we had to make.

\subsection{Deployment Considerations}
Before diving into the architectural details, it is important to recognize how the artifact must be deployed, as that impacts the architecture.
Acting as an OIDC Provider towards service clients, the bridge has full authority over transmitted user information with no accountability. Thus, the only feasible deployment option is for every service provider to deploy one themselves. Prior work has also reached this conclusion \cite{grunerAnalyzingInteroperabilityPortability2021}.

\subsection{Login Policy}

\begin{listing}
\scriptsize
\begin{minted}[frame=single,
               framesep=3mm,
               linenos=true,
               xleftmargin=13pt,
               tabsize=2]{js}
[{
  "credentialID": "expected_credential_for_email",
  "patterns": [{
    "issuer": "did:example:123",
    "claims": [{
        "claimPath": "$.credentialSubject.e_mail",
        "newPath": "$.email",
        "token": "id_token"
    }]
  },
  {
    "issuer": "did:example:456",
    "claims": [{
        "claimPath": "$.credentialSubject.email",
        "token": "id_token"
    }]
  }]
}]
\end{minted}
\caption{An example of a login policy for a service depending on the user's email address.} 
\label{json-example}
\end{listing}

To simplify setup, the whole system is configured with one main configuration file: the login policy. This defines what credentials to ask wallets for, which issuers to trust, what claims to accept, and how to transform those claims. While there are existing standards for specifying some of these, unifying everything into one file and format eliminates data duplication and avoids confusion about what files are security critical.

An example is shown in Listing \ref{json-example}. It defines a sign-in requesting one credential that contains a verified email address and is issued by one of two distinct issuers that organize this claim differently.
We will refer back to this login policy and discuss how exactly different components use it.

\subsection{Conceptual Components}
In this section, we describe our proposed system's conceptual architecture, as shown in Figure \ref{fig:architecture}. They represent units of functionality and do not necessarily need to be implemented separately.
The components are:

\subsubsection{OpenID Connect Provider}
An OpenID Connect Provider (OIDC Provider) is a service or software component that implements the OIDC protocol to enable secure and standardized user authentication and authorization. In our architecture, it is the interface for OIDC clients. At a minimum, it should support the common OAuth2.0 Authorization Code flow \cite{rfc6749}. The more robust and fully compliant this OIDC Provider is, the better the interoperability with existing OIDC clients.

It would ordinarily be possible for a client to request specific user data via OIDC scopes. For the user, that creates transparency about the transferred data. With our use of VCs, users actively choose what they share and give consent inside the wallet application when presenting VCs. That renders a consent step in the top-level OIDC flow unnecessary. Similarly, using OIDC scopes is superfluous because the Relying Party Service communicates what kind of data is expected to the wallet. This is feasible because the OIDC client and bridge are operated by the same party. We only expect clients to send the mandated default \texttt{openid} scope. Further scopes are ignored.

\subsubsection{Relying Party Service}
The Relying Party (RP) Service adheres to OID4VP with SIOPv2 to request and receive a Verifiable Presentation from a wallet. OID4VP dictates that a presentation definition from the DIF Presentation Exchange (PEX)\cite{identityPresentationExchange} specification is used to inform the wallet how many credentials of what kind are expected. This is not security critical in our system, but it is necessary to improve the user experience because wallets will usually pre-select suitable credentials based on the presentation definition.

\subsubsection{Presentation Definition Generator}
The Presentation Definition Generator creates a presentation definition based on a login policy, such as the one in Listing \ref{json-example}. The policy specifies one or more credentials it expects, giving each one a unique \texttt{credentialID}. For each expected credential, at least one pattern defines expected claims on a per-issuer basis. All patterns from the policy are turned into input descriptors requesting the presence of the claim without specifying filters. The Submission Requirement Feature\cite{identityPresentationExchange} is used to mirror that only one of the patterns from each expected credential needs to be fulfilled.
Should a service provider require more control, a custom array of input descriptors can be supplied optionally, overriding the Presentation Definition Generator.

\subsubsection{Verifiable Credential Verifier}
The Verifiable Credential Verifier's job is the cryptographic and structural verification of W3C VCs and VPs according to specification. In addition to that, revocation following the W3C's StatusList2021\cite{w3StatusList} proposal is supported. The status list server is optional and operated by the respective issuer that wishes to support revocation.

Our verifier also mandates that all VCs inside a VP are issued to the DID that the VP was signed by. This is a simple way to ensure a holder can only present VCs that make claims about him, and it is sometimes referred to as a holder binding.

\subsubsection{Policy Compliance Service}
To ensure that presented credentials contain the required claims and are issued by trusted issuers, the Policy Compliance Service evaluates a VP with respect to the login policy. An example policy is given in Listing \ref{json-example}. Each VC in a submitted VP needs to be matched to exactly one of the expected credentials in the policy. For each expected credential, the acceptable patterns are iterated, starting with the first one. Each pattern lists claims by JSONPaths following the Presentation Exchange specification\cite{identityPresentationExchange}. A VC matches a pattern if the issuer matches and if all required claims are present. By default, every claim is assumed to be required.
Our example policy expects one credential and declares two different patterns for it.

\subsubsection{Claim Processor}
Claims from the VP a user presents must be made available via OIDC in a way that is readable by current OIDC clients. The new \texttt{vp\_token} defined in OID4VP \cite{openidOpenIDVerifiable} could be used, but it is infeasible to rely on the quick adoption of a draft specification that will likely never be supported by many legacy systems depending on OIDC when we design for interoperability. Thus, claims need to be transferred into the established \texttt{id\_token} and \texttt{access\_token} \cite{openidFinalOpenID} to work with all OIDC clients. The Claim Processor's task is to turn a VP into token payloads according to a login policy.

Because existing services incorporating an OIDC client likely depend on standard OIDC claim names, transferring claims might not be sufficient, leading us to support renaming. This is also necessary if two trusted issuers use different claim names for the same piece of information, as can be seen in our example login policy in Listing \ref{json-example}. Every claim has a \texttt{claimPath}, which is a JSONPath. Optionally, \texttt{newPath} provides a new JSONPath for the claim value inside the corresponding OIDC token. It defaults to the last element in the \texttt{claimPath} and, if explicitly specified, must always point to exactly one location. The target token is specified using the \texttt{token} property, which defaults to the \texttt{access\_token} if not present. In summary, this allows arbitrary remapping of claim data.

In the case that a \texttt{claimPath} points to more than one value, a \texttt{newPath} is required. All claims in the original path are aggregated as a JSON object and indexed only by their ultimate JSONPath element. That object is written to the new path.

\subsection{DID Method Support}
The challenges posed by the DID standard's flexibility are numerous. The one mainly relevant for this work is the software\textemdash and perhaps protocol\textemdash support needed to resolve DIDs. We choose a pragmatic solution for this by recommending the use of simple lightweight DIDs, as defined in \cite{hoops2023taxonomy}. Every other DID method type would require another interface implementation custom to the used storage solution. Using \texttt{did:pkh}\cite{githubDidpkhdidpkhmethoddraftmdMain} specifically has the added benefit of leaving an upgrade path for blockchain-based DIDs. For example, a \texttt{did:pkh:eth} could also be used as the feature-richer \texttt{did:ethr}\cite{githubEthrdidresolverdocdidmethodspecmdMaster}.

\subsection{Limitations}
While we think that a bridge is a great option for fast prototyping and upgrading existing services, there are limitations to be aware of. First, services only have access to user attributes, not an interactive wallet connection. So if a service requires a wallet interaction to, for instance, obtain a user signature, relying on OIDC might not be advisable.


Next, we have chosen to keep our bridge as generic and simple as reasonably possible. As mentioned earlier, that entails support for a limited set of DID methods and for only one status list specification. However, future implementations could simply choose to support what is deemed appropriate for a given use case, provided expertise and development resources are sufficient. Some changes might be relatively easy.

Finally, the simple holder binding our bridge enforces makes it impossible for users with multiple VCs legitimately issued to different DIDs to present them in one VP and be accepted.

\section{Implementation}

\begin{figure*}[tbp]
\centering
\includegraphics[width=0.95\textwidth]{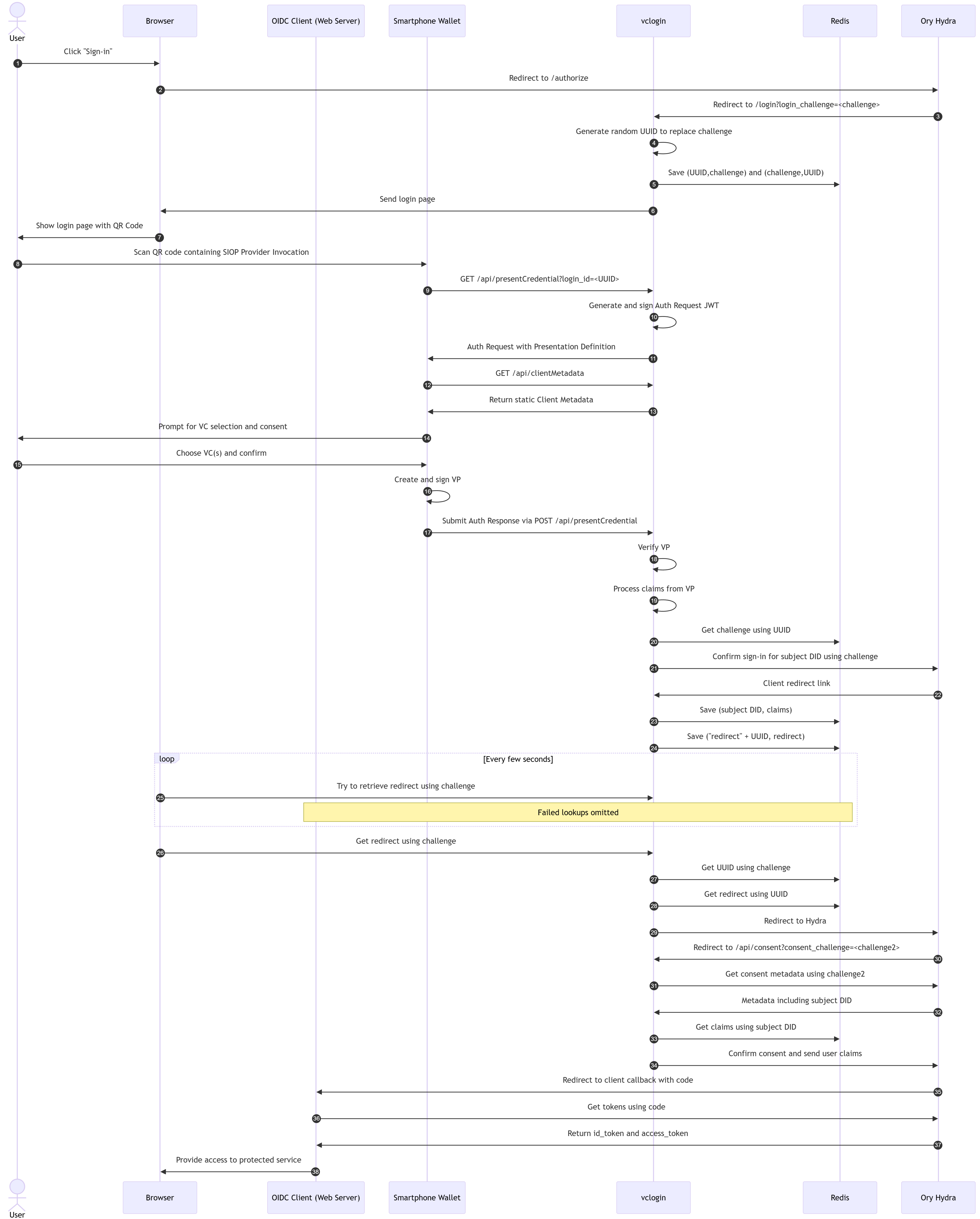}
    \caption{A simplified sign-in procedure using the authorization code flow with our bridge. At the start of the presented sequence, we assume that the user has accessed the web page provided by the OIDC client.}
    \label{fig:flow}
\end{figure*}

For our implementation, we were looking for a robust and configurable open-source IDP. We ended up choosing Ory Hydra\footnote{\url{https://github.com/ory/hydra}}, which is a free and open-source OIDC Provider. Its compliance is officially verified by the OpenID Foundation. Setting up Hydra is reasonably convenient via Docker\footnote{\url{https://www.docker.com/}} and the sign-in flow is by default incorporating redirects to custom web services for login and consent\footnote{\url{https://www.ory.sh/docs/oauth2-oidc/custom-login-consent/flow}}. That lets us build our proof of concept without touching Hydra's source code.

Our custom implementation called \emph{vclogin} will take on the role of the login and consent services for Hydra. Additionally, we bundle all other remaining conceptual components into vclogin to keep deployment complexity down.
For this reason, we chose to build with Next.js\footnote{\url{https://nextjs.org/}}, as it allows us to combine user-facing web pages with API routes in a single code base.

Our code base comes with a Docker Compose file that contains all the configuration to run the entire bridge. That includes a PostgreSQL\footnote{\url{https://www.postgresql.org/}} Database for Hydra and a Redis\footnote{\url{https://redis.io/}} for vclogin. The required configuration is minimal and documented. The code is intended for prototyping and as a reference. It has not undergone a security audit, and it is not intended to be used in production.

The Redis store only acts as temporary storage, and all data is set to expire within minutes. For our use case, the PostgreSQL database can also be reset without effect. Thus, the entire bridge is essentially stateless and can be migrated without a data and database migration.

\subsection{Detailed Protocol Flow}

The biggest challenge in engineering the bridge lies in the specifics of the cross-device sign-in flow. As an example, we will now examine the adaption of perhaps the most common OAuth 2.0 flow: the Authorization Code flow\cite{rfc6749}. Our protocol flow can be seen in Figure \ref{fig:flow}. We have slightly simplified some parts while keeping the depiction close to the technical reality. For example, we have omitted responses for Redis interactions.

\begin{listing}
\scriptsize
\begin{minted}[frame=single,
               framesep=3mm,
               linenos=true,
               xleftmargin=17pt,
               tabsize=2]{js}
openid-vc://
?client_id=<BRIDGE_DID>
?request_uri=<BACKEND_URL>/api/presentCredential
%3Flogin_id=<UUID>
\end{minted}
\caption{QR code contents for starting a credential exchange process with a wallet via SIOPv2 and OID4VP.} 
\label{lst:openid-vc}
\end{listing}

At the start of the depicted flow, the user is assumed to have opened a service website hosted by the OIDC client web server in their browser. The flow starts normally until the user is redirected to the login page. All it features is a QR code that contains the necessary information to involve the smartphone wallet in the flow, which is depicted in Listing \ref{lst:openid-vc}.

The \texttt{request\_uri} tells the wallet where to fetch a presentation request. Notably, we add a custom query parameter called \texttt{login\_id} to use as a challenge for the later presentation exchange and to allow the vclogin backend to know which browser and wallet are involved in the same sign-in procedure. Generating a UUID as the \texttt{login\_id} in step 4 is necessary because the login challenge generated by Hydra is so large it cannot fit into a readable QR code. To keep track of this mapping between UUID and challenge, it is written to Redis. When the wallet asks for a presentation request in step 9, it also automatically provides vclogin with the \texttt{login\_id}. It is used as the VP challenge during the following standardized Presentation Exchange flow.

After step 17, vclogin has received a VP. Following successful processing, vclogin exchanges the UUID challenge from the VP back into the login challenge to confirm the authentication and authorization to Hydra. The subject is identified by its used DID. Additionally, all processed claims from the VP are written to Redis to be accessible during the next phase.

At this point, the active role in the sign-in process needs to be handed back to the browser. To do so, vclogin has written the \texttt{redirect\_uri} for the browser to Redis. Using the original login challenge, the browser periodically requests the redirect to see if the VP submission has happened yet. When it succeeds, it executes the redirect in step 29.

Now, the consent phase of the authorization code flow would take place, but as we do not need to ask for consent a second time, vclogin automatically completes it. Using the new \texttt{consent\_challenge}, vclogin can request metadata on this consent process from Hydra in step 31. Most importantly, that includes the subject DID, which is used to retrieve the processed claims that were previously written to Redis in step 23. Then, vclogin confirms the user's consent to Hydra and provides the claim data for tokens in step 34.

Finally, the sign-in procedure concludes as normal for an authorization code flow from step 35 and on. The only noteworthy exception is that the bridge never provides a \texttt{refresh\_token} because the bridge must not depend on caching critical identity data and should remain stateless in the larger picture.

\section{Evaluation}
We evaluated our design and implementation in two ways: practically by performing a realistic sign-in and conceptually by recalling the open challenges we identified in the beginning.

\subsection{Practical Testing}

\begin{listing}[t]
\scriptsize
\begin{minted}[frame=single,
               framesep=3mm,
               linenos=true,
               xleftmargin=17pt,
               tabsize=2]{js}
"credentialSubject": {
  "id": "did:key:z6M...",
  "email": "name@example.com",
  "type": "EmailPass",
  "issuedBy": { "name": "Altme" }
}
\end{minted}
\caption{The \texttt{credentialSubject} key in an Altme Proof of email VC with shortened \texttt{id} field.} 
\label{lst:email_credentialSubject}
\end{listing}

\begin{listing}[t]
\scriptsize
\begin{minted}[frame=single,
               framesep=3mm,
               linenos=true,
               xleftmargin=17pt,
               tabsize=2]{bash}
DID_KEY_JWK=<jwk>
EXTERNAL_URL=https://examplebridge.com
LOGIN_POLICY=./acceptEmailFromAltme.json
PEX_DESCRIPTOR_OVERRIDE=./descrEmailFromAltme.json
\end{minted}
\caption{An example \texttt{.env} file used to configure the SSI-to-OIDC bridge with a JWK placeholder.} 
\label{lst:env}
\end{listing}

\begin{listing}[t]
\scriptsize
\begin{minted}[frame=single,
               framesep=3mm,
               linenos=true,
               xleftmargin=17pt,
               tabsize=2]{js}
[{
  "credentialID": "one",
  "patterns": [{
    "issuer": "did:web:app.altme.io:issuer",
    "claims": [{
      "claimPath": "$.credentialSubject.email",
      "token": "id_token"
    }]
  }]
}]
\end{minted}
\caption{A login policy accepting a VC issued by Altme, containing an email.} 
\label{lst:emailPolicy}
\end{listing}

\begin{listing}[t]
\scriptsize
\begin{minted}[frame=single,
               framesep=3mm,
               linenos=true,
               xleftmargin=17pt,
               tabsize=2]{js}
{
  ...
  "email": "name@example.com",
  "sub": "did:key:z6M..."
}
\end{minted}
\caption{A shortened \texttt{id\_token} payload containing an email.} 
\label{lst:idToken}
\end{listing}

For real-world testing, we used the Altme Wallet\footnote{\url{https://github.com/TalaoDAO/AltMe}}. It is open-source, actively worked on, and regularly updated to its newest version on Apple's App Store and Google's Play Store for convenient installation. This particular wallet combines the capabilities of a crypto wallet and an SSI wallet. For our testing, it is essential that the wallet supports OID4VP with SIOPv2. In Altme, a \emph{Proof of email} VC is offered, which is one of the VCs we used in our tests. The relevant claims in that VC are shown in Listing \ref{lst:email_credentialSubject}. It is signed by a \texttt{did:web} and contains a StatusList2021 entry.

As an OIDC client, we will use the client Ory includes in Hydra's command line interface (CLI) for testing\footnote{\url{https://www.ory.sh/docs/cli/ory-perform-authorization-code}}. It hosts a website with a sign-in button and shows session metadata upon successful sign-in.

Before starting the SSI-to-OIDC bridge, we need to ensure a fitting configuration. The environment variables seen in Listing \ref{lst:env} need to be set, where \texttt{DID\_KEY\_JWK} is the key used as a \texttt{did:key} to authenticate to the wallet with. While \texttt{PEX\_DESCRIPTOR\_OVERRIDE} is optional, we use it here to combat a crash likely stemming from Altme's implementation of the Presentation Exchange specification by only requesting the credential type. The login policy we used can be found in Listing \ref{lst:emailPolicy}.

After starting the bridge, we can use Hydra's CLI to register a new client and then start our test client. Now, we perform these steps:
\begin{enumerate}
    \item Click "Authorize application" and get redirected.
    \item Scan the QR code below the text "Scan the code to sign in!" with the wallet.
    \item Select one email VC from the list and press "Present" in the wallet.
    \item Confirm the choice with the biometrics of the smartphone.
    \item Website redirects to a page showing the received tokens, with the \texttt{id\_token} being as configured. It can be seen in Listing \ref{lst:idToken}.
\end{enumerate}


\subsection{Addressed Challenges}

Looking at all of the overall challenges we identified, we have addressed all of them in various ways:

\begin{itemize}
\setlength{\itemindent}{-1em}
    \item[]\textbf{Fragmented Presentation Protocols} Our bridge relies fully on SIOPv2 and OID4VP, which show considerable promise to become the quasi-default, learning from and being associated with established IAM.
    \item[]\textbf{SSI Ecosystem Dependence} Until further standards are established, we avoid custom protocols, stick to simple lightweight DID methods, and support revocation via just the established StatusList2021 spec.
    \item[]\textbf{High Bridge Complexity} Our design has one configuration file with powerful defaults. A local test instance of the bridge can be run within minutes.
    \item[]\textbf{Lack of Concrete Technical References} We have contributed reference code and an in-depth look at a bridged authorization code flow.
\end{itemize}






\section{Conclusion}
We have proposed an architecture that can simplify the adoption of SSI for sign-ins and described it in detail. To prove its feasibility, we have fully implemented this SSI-to-OIDC bridge and successfully tested it with existing software and realistic hardware.
Our complete code repository is available on GitHub\footnote{\url{https://github.com/GAIA-X4PLC-AAD/ssi-to-oidc-bridge}} and will be further evaluated as part of a Gaia-X project for decentralized business-to-business cooperation.

In the long term, a mechanism to standardize VC types and claim names is needed. Otherwise, any bridge needs to remap claims as soon as more than one issuer is involved, which is almost inevitable, given that SSI has decentralization at its core. Remapping claims complicates configuration and could introduce security-critical errors. This is a governance issue but also a technology deficit: relying on VC contexts and their types is not secure. Similarly, an easier way to organize collections of trusted issuers would simplify the configuration and administration of a bridge.

\section*{Acknowledgment}
This work has been funded by the German Federal Ministry of Economic Affairs and Climate Action (BMWK) under grant 19S21006N. The responsibility for the content of this publication lies with the authors.

\bibliographystyle{IEEEtran}
\bibliography{bibliography}

\end{document}